\documentclass[fleqn,usenatbib]{mnras}

\usepackage{newtxtext,newtxmath}
\usepackage{float} 

\usepackage[T1]{fontenc}

\DeclareRobustCommand{\VAN}[3]{#2}
\let\VANthebibliography\thebibliography
\def\thebibliography{\DeclareRobustCommand{\VAN}[3]{##3}\VANthebibliography}

\usepackage{graphicx}	
\usepackage{amsmath}	
\usepackage{longtable,rotating} 
\usepackage{caption}
\usepackage{import}
\usepackage{hyperref}
\usepackage{endnotes}
\usepackage{siunitx} 
\usepackage[normalem]{ulem}

\newcommand{\HI}{\mbox{H{\sc i}}}
\newcommand{\kms}{\mbox{km\,s$^{-1}$\,}} 
\newcommand{\Msun}{M$_\odot$} 
\newcommand{\MHI}{$M_{\rm {HI}}$}
\newcommand{\MHIunit}{log$(M_{\rm HI}/$M$_\odot)$}
\newcommand{\vhel}{\mbox{$V_{\rm {hel}}$}\,}

\newcommand{\Wfifty}{\mbox{${W}_{\mathrm50}$}}
\newcommand{\Wtwenty}{\mbox{${W}_{\mathrm20}$}}

\title[SARAO MeerKAT GPS -- the Great Attractor Connection]{H{\selectfont \textsc i}\ Galaxy Signatures in the SARAO MeerKAT Galactic Plane Survey -- I. Probing the richness of the Great Attractor Wall across the inner Zone of Avoidance}

\author[N. Steyn et al.]{
Nadia Steyn$^{1,2}$,\thanks{E-mail: nadia.steyn@icrar.org}
Ren{\'e}e C. Kraan-Korteweg$^{1}$,
Sambatriniaina H. A. Rajohnson$^{1}$,
Sushma Kurapati$^{1}$,
\newauthor 
Hao Chen$^{1,3}$,
Bradley Frank$^{4,5,6,1}$, 
Paolo Serra${^7}$,
Lister Staveley-Smith$^{2,8}$,
Fernando Camilo$^{5}$,
\newauthor 
and Sharmila Goedhart$^{5,9}$
\\
$^{1}$ Department of Astronomy, University of Cape Town, Private Bag X3, Rondebosch 7701, South Africa\\
$^{2}$ International Centre for Radio Astronomy Research (ICRAR), The University of Western Australia, 35 Stirling Highway\\
$^{3}$ Research Center for Intelligent Computing Platforms, Zhejiang Laboratory, Hangzhou 311100, China\\
$^{4}$ UK Astronomy Technology Centre, Royal Observatory Edinburgh, Blackford Hill, Edinburgh EH9 3HJ, UK\\
$^{5}$ South African Radio Astronomy Observatory (SARAO), 2 Fir Street, Observatory, 7925, South Africa\\
$^{6}$ The Inter-University Institute for Data Intensive Astronomy (IDIA), and University of Cape Town, Private Bag X3, Rondebosch, 7701, South Africa\\
$^{7}$ INAF -- Osservatorio Astronomico di Cagliari, Via della Scienza 5, 09047, Selargius, CA, Italy\\
$^{8}$ ARC Centre of Excellence for All Sky Astrophysics in 3 Dimensions (ASTRO 3D), Australia\\
$^{9}$ SKAO, 2 Fir Street, Black River Park, Second Floor, Block A, Cape Town, 7925\\
}

\date{Accepted XXX. Received YYY; in original form ZZZ}

\pubyear{2023}

\begin{document}
\label{firstpage}
\pagerange{\pageref{firstpage}--\pageref{lastpage}}
\maketitle

\begin{abstract}
This paper presents the first \HI\ results extracted from the SARAO MeerKAT Galactic Plane Survey (SMGPS) -- a narrow strip ($\Delta b \sim 3\degr$) along the southern Milky Way. The primary goal consisted in tracing the Great Attractor (GA) Wall across the innermost Zone of Avoidance. We reduced a segment spanning the longitude range $302\degr \leq \ell \leq 332\degr$ for the redshift range $z \leq 0.08$. The superb SMGPS sensitivity (rms\,=\,$0.3-0.5$ mJy\,beam$^{-1}$ per 44\,\kms~channel) and angular resolution ($\sim$\ang{;;31}$\times$ \ang{;;26}) lead to a detection limit of \MHIunit $\geq$ 8.5 at the GA distance ($\vhel \sim 3500 - 6500$\,\kms).
A total of 477 galaxy candidates were identified over the full redshift range. A comparison of the few \HI\ detections with counterparts in the literature (mostly HIZOA) found the \HI\ fluxes and other \HI\ parameters to be highly consistent. 
The continuation of the GA Wall is confirmed through a prominent overdensity of $N = 214$ detections in the GA distance range. 
At higher latitudes, the wall moves to higher redshifts, supportive of a possible link with the Ophiuchus cluster located behind the Galactic Bulge. This deep interferometric \HI\ survey demonstrates the power of the SMGPS in improving our insight of large-scale structures at these extremely low latitudes, despite the high obscuration and continuum background.

\end{abstract}



\section{Introduction}

The so-called Zone of Avoidance (ZoA) bisects a number of large-scale structures that are relevant for derivations of the motion of the Local Group and bulk flows, e.g., the Local Void \citep[LV; e.g.,][]{rkk2008,tully2019}, the Vela Supercluster \citep[SCL;][]{RKK2017}, and the Great Attractor (GA). The GA was originally discovered from one of the earliest whole-sky peculiar velocity data-sets \citep{dressler1987,Lynden-Bell1988}. Matching the GA to an overdensity in galaxies remained a challenge, however, because of its location behind the Milky Way. Dedicated galaxy searches and redshift follow-ups close to the ZoA \citep{Woudt&RKK2001,woudt2004,Radburn-Smith2006} suggested the GA to consist of a large wall-like structure with the rich Norma cluster at its core \citep[ACO\,3627; $\ell,b, V_{\rm hel} = 325\fdg3,-7\fdg2, 4871$\,\kms;][]{RKK1996,woudt2004}. But these could not penetrate the latitudes below $|b|\la 5\degr$, and the prominence and shape of the Wall across the inner ZoA was never fully ascertained. Only systematic \HI\ surveys prevail at these latitudes, since they are unaffected by obscuration -- as first demonstrated by \cite{Henning1992}. This was one of the reasons the Parkes \HI\ ZoA survey was conceived \citep[$212\degr\leq\ell\leq36\degr, b\leq\pm5\degr$; $V_{\rm hel}< 12700\,\kms$;][henceforth~HIZOA]{Staveley-Smith2016}. With a 15\farcm5 beam size and 6 mJy\,beam$^{-1}$ chan$^{-1}$ noise (rms), it was optimised to uncover normal spirals at the GA distance range. And indeed HIZOA succeeded in tracing the GA across the optically opaque part of the ZoA, albeit with the low detection rate of $\sim$1\,galaxy per 2\,deg$^2$ on average. However, the single dish data became increasingly incomplete where the continuum background rises above T$_B > 7$K ($|b| \la 1\fdg5$), reaching zero completeness for T$_B > 16$K.

\begin{figure*} 
    \centering
    \includegraphics[width=1\textwidth]{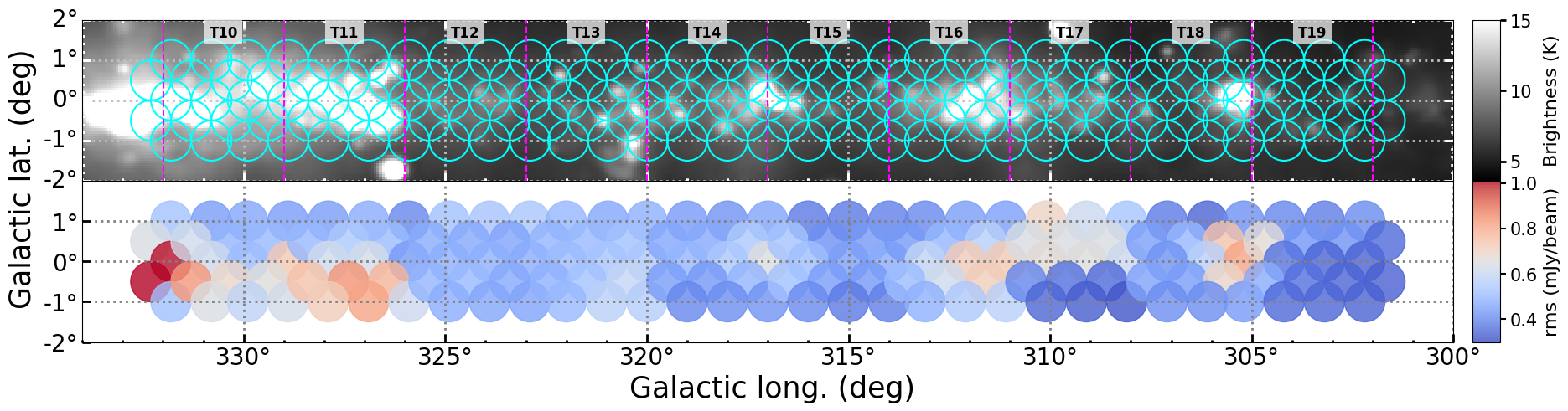}
    \caption{Top: The 1.4\,GHz radio continuum image based on CHIPASS \citep{Calabretta2014}. The positions of the GA-SMGPS pointings are overlaid in cyan with radii of $0\fdg5$ (FWHM at $z$=0). The boundaries separating the mosaics are approximately traced in magenta, with corresponding labels for our mosaic names (explained in text). Bottom: the sensitivity of the GA-SMGPS fields, with colours representing the measured rms of a central plane in the respective field.} 
    \label{fig:rms_circles}
\end{figure*}

The SARAO MeerKAT Galactic Plane Survey (SMGPS; Goedhart et al., in prep) is ideally suited to bridge this gap. The interferometric data will be less affected by continuum residuals due to spatial filtering.
Moreover, the typically $\sim$1~hr on-source integration per pointing leads to a sensitivity of 0.3--0.5 mJy\,beam$^{-1}$ per 44\,\kms\,channel -- an order of magnitude better than HIZOA\,--\,while the spatial resolution of $\sim$\ang{;;31}$\times$ \ang{;;26} (15\farcm5 in HIZOA) will resolve all galaxies with \HI-masses above $10^8$\,M$_{\odot}$ at a typical distance of 70 Mpc, according to the \HI\ mass-size relation of \cite{wang2016}.

This first of a series of papers analysing the SMGPS \HI\ data will zoom in on the GA-Wall ZoA-crossing. Here we cover the longitude range $302\degr\leq\ell\leq332\degr$, where the higher longitude side was selected to find signatures for the GA Wall extension emanating from the Norma cluster \citep[e.g.,][]{woudt2008}, while the other end will encompass the surroundings of the galaxy cluster centered on PKS\,1343--601 \citep{Nagayama2004, schroeder2007}.

\section{Observations}
\label{section:GPS_observations}

\subsection{Data Reduction and Imaging}
The SMGPS observations were carried out between July 2018 and March 2020 with the MeerKAT L-band receiver and the SKARAB-4k correlator (Goedhart et al., in prep). The 4096 channels across the frequency range 856 -- 1712 MHz led to a frequency resolution of 209\,kHz, producing a velocity resolution of 44.1~km\,s$^{-1}$ at $z=0$. 

For this study, we reduced a subsection of the full 528~deg$^2$ SMGPS: a $30\degr \times 3 \degr$ survey area comprising 157 pointings arranged in a closely-spaced grid (henceforth the GA-SMGPS pointings). The calibration, imaging and mosaicking were completed on the \href{www.Ilifu.ac.za}{Ilifu} cloud computing facility, hosted by the Inter-University Institute for Data Intensive Astronomy (\href{www.idia.ac.za}{IDIA}).
We processed the mostly RFI-free frequency range $1308 - 1430$\,MHz, using both orthogonal polarizations, but excluded frequencies dominated by gas emission from the Milky Way ($1419.46 - 1421.35$\,MHz). This allows us to investigate the large-scale structure out to $\vhel < 25000$~\kms; far enough to include the volume relevant to (residual) bulk flow studies \citep[$\vhel > 16000\,\kms$;][]{Springob16,Scrimgeour16}, and achieving excellent sensitivity well into the dwarf regime at the distance of the GA overdensity ($\sim3500-6500$\,\kms).

Our visibility data were reduced with \href{https://caracal.readthedocs.io/}{CARACal}, 
an automated end-to-end pipeline for radio interferometry data \citep{Jozsa2020} which invokes \sffamily\textsc{stimela} \normalfont \,--\,a Python package based on container technology \citep{makhathini2018}.
The reduction process includes flagging, cross-calibration, self-calibration, Doppler-tracking correction, and continuum subtraction (see Rajohnson et al., in prep. for details). 

The latter is particularly important when working in the innermost ZoA, where continuum sources are abundant. This is seen in the top panel of Fig.~\ref{fig:rms_circles}, which displays the 1.4~GHz background continuum temperature, as derived from CHIPASS by \cite{Calabretta2014}, extracted here for the GA-SMGPS region. Overlaid in cyan are the positions of the 157 GA-SMGPS pointings with their FWHM radius of $0\fdg5$ (at $z$=0). The bottom panel shows the colour-coded rms per individual field as determined from the central plane of the respective \HI-data cube (<rms> = 0.51~mJy). 
The slight overall rise in the amplitude of the rms towards higher longitudes is due to an increase in the continuum emission of our Galaxy, while the patches of higher rms show a clear trend with bright emission sources. These are primarily caused by extended diffuse continuum sources. They raise the receiver temperature locally and may reduce the \HI\ detectability.

The \HI~data were imaged with \sffamily\textsc{wsclean}\normalfont, using a pixel scale of $\ang{;;3}$ and a Briggs robust weighting of $r=0$. We employed a UV-tapering of $\ang{;;15}$ because the profiles of the non-tapered point spread functions (PSFs, or `dirty beams') revealed significant wings, causing clear deviations from Gaussianity, which would make it difficult to calculate integrated flux. 
The taper increased the beam size from a mean of $\sim$\ang{;;11}$\times$ \ang{;;9} to about $\sim$\ang{;;31}$\times$ \ang{;;26}, which  is sufficient for our science goals.

\subsection{Mosaicking} 
\label{section:mosaicking}

\begin{figure}
    \includegraphics[width=0.2525\textwidth]{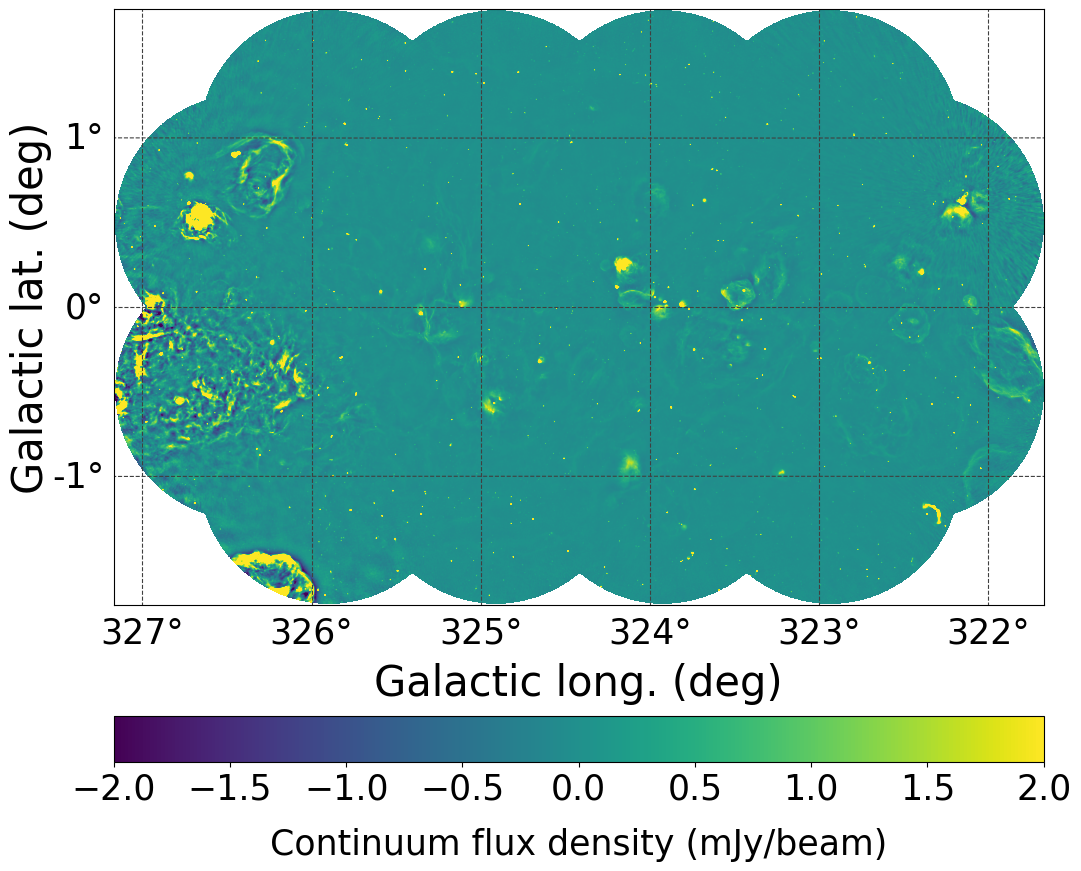} 
    \hspace*{-0.18cm}
    \includegraphics[width=0.23\textwidth]{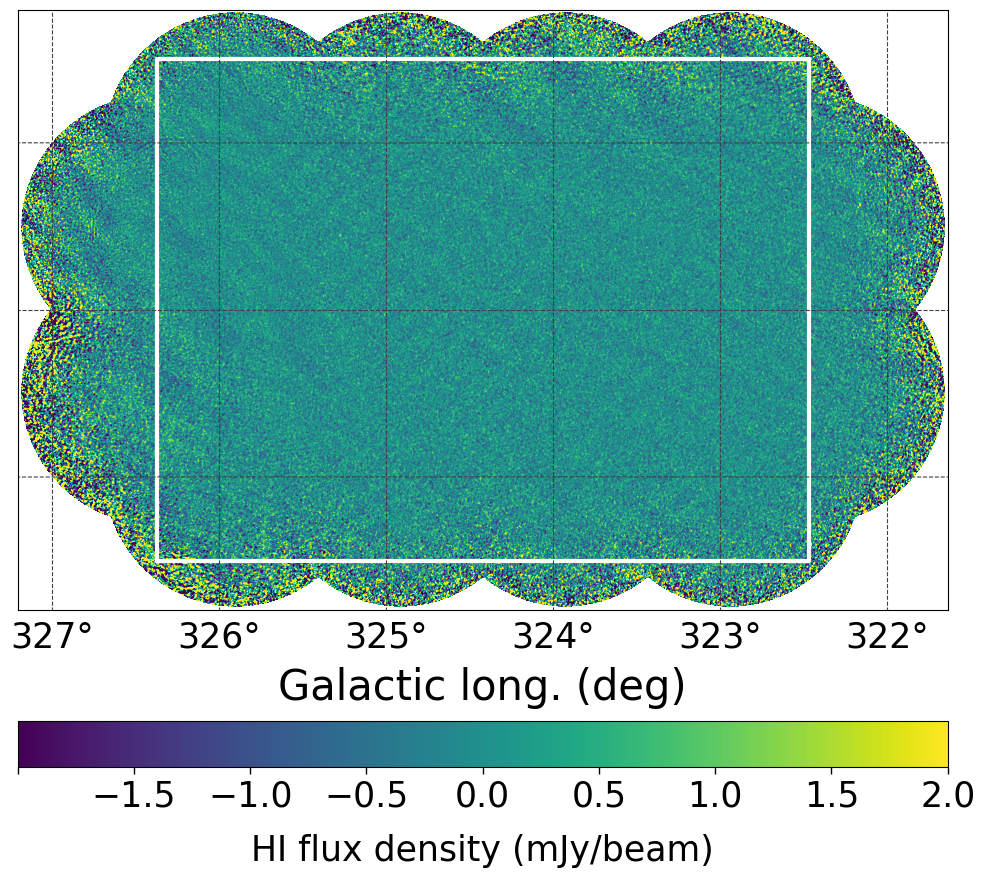}    
    \caption{Continuum and \HI~channel map of mosaic T12. The colour scale (linear stretch) indicates the flux variations. While the continuum mosaic reveals both compact and diffuse continuum sources, the centrally located \HI\ plane is\,--\,apart from the borders\,--\,fairly uniform (rms = 0.42 mJy.}
    \label{fig:T12_mosaic}
\end{figure}

The final \HI\ fields were imaged with $0.8$° radii (corresponding to a cut-off at 20\% sensitivity) and were primary-beam corrected using the method described in \cite{Mauch2020}. These were combined within CARACal into ten consecutive, overlapping \HI~mosaics, consisting of 22 pointings per mosaic. 
They are designated as T10 to T19; where "T" is for "tile", in accordance with SMGPS observation nomenclature. 
Each mosaic has an on-sky footprint of roughly $5\degr\times\,3\fdg5$ (see Fig.~\ref{fig:T12_mosaic}) with a generous overlap of $\Delta\ell\sim2\fdg5$. The continuum data were likewise imaged with a pixel size of $\ang{;;3}$ and constructed into two-dimensional mosaics.
The central $\Delta\ell\sim3\fdg5$ of each \HI\ mosaic was used for analysis\,--\,a region with near-uniform sensitivity (apart from artefacts) given the hexagonal sampling pattern and small offsets between pointings. 

Figure~\ref{fig:T12_mosaic} shows the continuum mosaic (left) next to a central \HI\ plane from mosaic T12. Inspecting the \HI\ images for continuum residuals, we found that compact continuum sources overall are well subtracted, but large-scale diffuse continuum sources caused an increase in the noise of the output image (e.g., around $\ell \gtrsim 325^{\circ})$.

The measured rms of the ten mosaics range from 0.39--0.53 mJy (mean mosaic rms\,=\,0.47 mJy). This includes the overall increase in $T_{\rm B}$ towards higher longitudes {\em and} local variations induced by continuum residuals -- the latter of which could hamper source extraction. 
Apart from some remaining Global Positioning Satellites RFI spikes ($1376 - 1386$\,MHz), the spectral baselines within the \HI\ mosaics mostly remain flat out to the edge of the survey ($\vhel <25000$\,\kms), with little variation in the noise level.

\section{Source Finding}
\label{section:GPS_sofia}

Given the size of mosaics ($\sim$$6600 \times 4300 \times 570$ pixels per $\sim$65\,GB cube), visual source finding becomes prohibitive, and automated source-detection algorithms, such as SoFiA used in this work \citep{Serra2015,Westmeier2021}, become essential. On the other hand, having an existing source list to optimise the parameter settings is incredibly advantageous. We therefore searched one full mosaic visually using CARTA \citep{CARTA2021}\,--\,a remote image visualisation and analysis tool. We chose T12 as the representative mosaic for the visual search. It lies close to the Wall that seems to be emanating from the Norma Cluster ($\ell\sim\,$325°), moreover, its relatively high number of HIZOA detections (12) increases the likelihood of uncovering a fair number of SMGPS \HI\ detections for the optimisation process\,--\,and indeed 90 sources were found in T12 alone. The resulting visual catalog was used as a template to optimise the SoFiA~v2.3.1 parameter settings.

We chose a central rectangular search area ($3\fdg5 \times 3\fdg0$) within which to execute SoFiA (see white rectangle in Fig.~\ref{fig:T12_mosaic}). This was selected to be slightly larger than the longitudinal offset between the mosaics ($\Delta \ell = 3\degr$) to not miss galaxies at the edges, and to allow for an internal consistency check of source identification and parameter extraction.
We cross-matched the visual catalog of T12 to the various outputs from different SoFiA runs, where it became apparent that we required more than one set of parameters to capture the maximum amount of visual detections. 
This led to an optimised SoFiA source-finding strategy consisting of three consecutive SoFiA runs: in the first run the maximum z-length \sffamily(\textsc{Linker.maxSizeZ}) \normalfont was set to 21, and was increased in a second run to 31, while the flux threshold (\sffamily\textsc{scfind.threshold}) \normalfont was changed from 4.0$\sigma$ to 3.5$\sigma$ in the third run. The outputs were merged and false positives removed\,--\,being especially critical of detections extracted from noisy regions. Our final parameter settings are available in \hyperref[appendixa]{Appendix A}, and further details of the optimisation process can be found in \cite{Steyn2023}. 

The above-described strategy was implemented on all ten mosaics. The resulting detection lists were adjudicated by more than one member of the team, and combined into one final catalog with the duplicates removed. The adjudication was based on an assessment of the moment maps (total intensity and velocity), the global \HI\ spectrum and its S/N ratio, and also the \HI\ extent in kpc with respect to its redshift and observed \HI-rotation. All sources are resolved by minimally two beams.

\section{Results}
\label{section:GPS_results}

\subsection{The GA-SMGPS Catalog} 
\label{section:GPS_catalog}

Here we discuss the final \HI-catalog and the quality of the extracted \HI\ parameters. We catalogued a total of 477 galaxy candidates in the GA-SMGPS region: 405 solid detections (marked in the catalog as flag 1) and 72 with slightly lower confidence (flag 2). 
Flag two detections are still reasonably strong detections but in higher noise areas or with low S/N, and all candidates are used in analysis. 
The GA-SMGPS galaxy catalog, given in \hyperref[appendixb]{Appendix B}, lists the ID as SMGPS-HI Jhhmmss-ddmmss, the mosaic name, the Galactic coordinates $\ell$, $b$, the \HI\ parameters $S_{\text{peak}}$, $S_{\text{int}}$, err$_{S_{\text{int}}}$, rms, $V_{\text{hel}}$, \Wtwenty, \Wfifty, log$M_{\rm HI}$, and lastly, flag and note/counterpart. Details are explained in \hyperref[appendixb]{Appendix B}. An atlas of all detections (moment maps and \HI-profiles) is also available online. 

Flux verification and reproducibility is a fundamental step in assessing the data quality of \HI~images. At these low latitudes, no optical counterparts are known and the primary source that allows for a comparison is HIZOA. 
A few infrared counterparts identified in literature (three mentioned in HIZOA and six new) are given in the last column of our galaxy catalog (\hyperref[appendixb]{Appendix B}). 
For comparison, the survey parameters of GA-SMGPS, HIZOA and the forthcoming WALLABY survey \citep{koribalski2020} are summarised in Table~\ref{tbl:gps-hizoa}.

\begin{table*}
    \captionsetup{justification=centering} 
    \caption{\centering Survey parameters of GA-SMGPS, HIZOA {\bf and WALLABY}.} 
    \label{tbl:gps-hizoa}
    \begin{small}
    \centering
    \begin{tabular}{lccc}
    \hline
    Parameter & GA-SMGPS & HIZOA & {\bf WALLABY}\textsuperscript{\textdagger} \\ \hline
    Telescope & MeerKAT & Parkes & ASKAP \\
    Date of observations & 2018 -- 2019 & 1997 -- 2000 & 2022 -- ongoing \\
    Sky coverage & 302°$\leq\ell\leq$ 332°;~$|b|$$\leq$1$\fdg$5 & 212°$\leq\ell\leq$36°;~$|b| $$\leq$5$\degr$ & $-90\degr \leq \delta \leq +30\degr$ \\
    Velocity range ($cz$) & <\,25000\,\kms &  <\,12740\,\kms & <\,77000\,\kms\\
    Velocity resolution & 44.1\,\kms\ & 27\,\kms$^*$ & 4\,\kms\ \\ 
    Angular resolution & $\sim(\ang{;;31} \times \ang{;;26}$) & $15\farcm5$ & $\sim\ang{;;30}$ \\
    Integration time pointing$^{-1}$ & $\sim 3600$\,s & 2100\,s & 2$\times$8\,hr \\ 
    Measured rms (per channel) & 0.3--0.5 mJy\,beam$^{-1}$ & 6 mJy\,beam$^{-1}$ & 1.6 mJy\,beam$^{-1}$ \\ \hline
    \parbox{3cm}{*\footnotesize{after Hanning smoothing}} & \multicolumn{3}{r}{\parbox{8cm}{\footnotesize{\textsuperscript{\textdagger}Widefield ASKAP L-band Legacy All-sky Blind surveY}}} \\
    \end{tabular}
    \end{small}
\end{table*}

Forty (out of 42) HIZOA galaxies in the GA-SMGPS data were recovered by SoFiA. 
The HIZOA galaxy J1532-56 at $V \sim 1400$\,\kms, identified as a face-on, very extended, ring-like disk in the GA-SMGPS, has no entry in our catalog: the flux per voxel was too low to allow for a robust parameterisation. Earlier observations of this source with the Australian Telescope Compact Array \citep[ATCA;][]{staveley-smith1998} reported on its exceptionally low column density and large \HI-disk.
Interestingly, the SMGPS data finds the \HI-disk to be even more extended than the ATCA FoV. We performed deeper observations with the MeerKAT 32k correlator for further investigations of the properties of this particular source, and results will be presented in a later study. HIZOA source J1534-56A could not be parametrised for the same reason (low column density).

\begin{figure} 
    \centering 
    \includegraphics[width=0.35\textwidth]{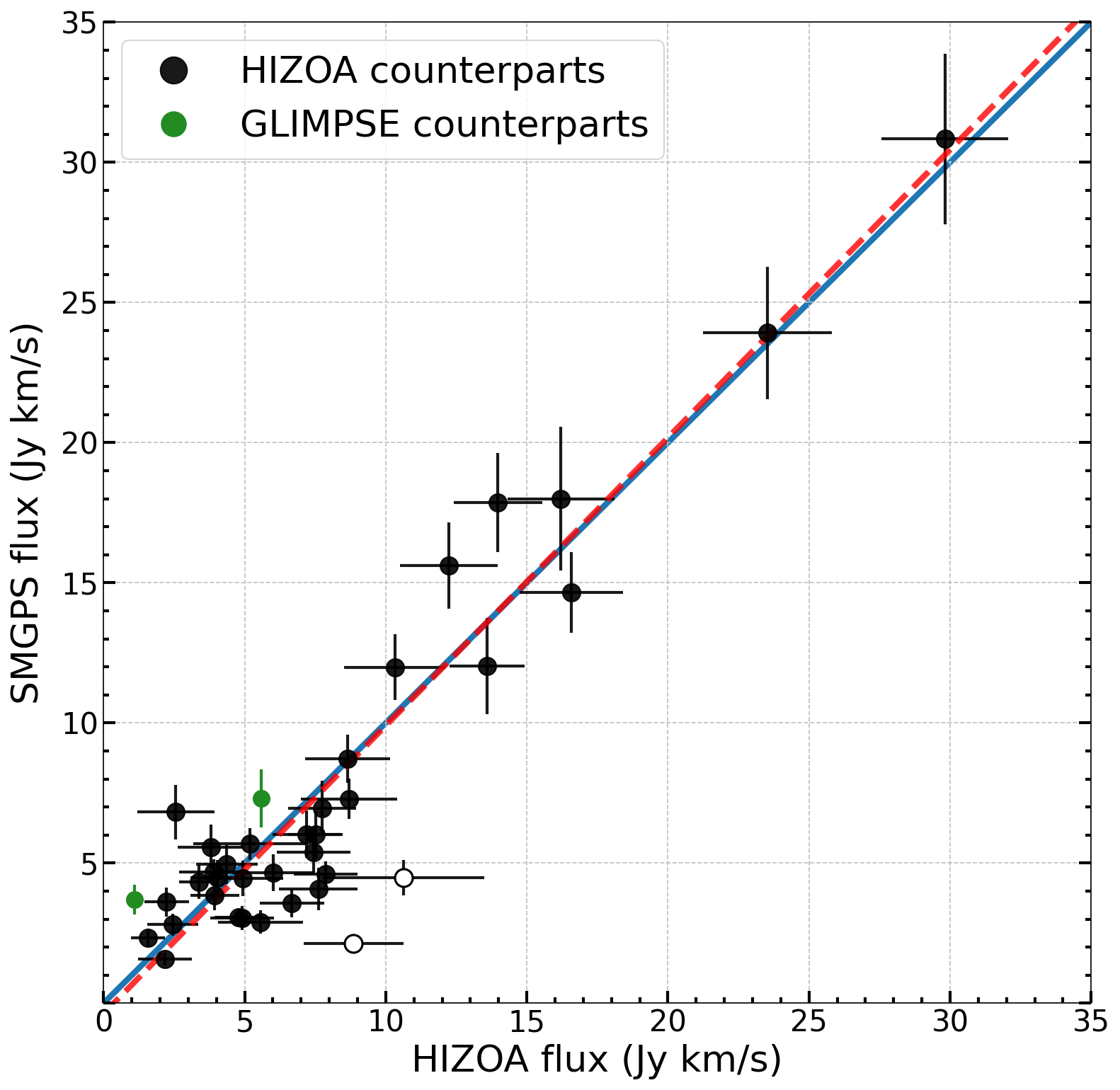}
    \caption{Comparison between the integrated fluxes ($S_{\rm int}$) of GA-SMGPS detections with known counterparts: 36 HIZOA sources and 2 ATCA detections for GLIMPSE galaxies (green). The blue solid line represents a one-to-one relation, the red dashed line a linear fit. Two outliers  were excluded from the fit because their HIZOA baselines are unreliable (open circles).} 
    \label{fig:flux_comparison}
\end{figure}

Of the 40 HIZOA sources we recovered, four were resolved into compact galaxy pairs/groups; not surprising given the large beam of the HIZOA survey. 
The remaining 36 integrated \HI-flux values ($S_{\text{int}}$) are plotted in Fig.~\ref{fig:flux_comparison} versus their SMGPS counterparts (black markers). The two additional points (green markers) are based on 12~hr synthesis \HI\ observations with ATCA of galaxies identified in GLIMPSE \citep[Galactic Legacy Infrared Mid-Plane Survey Extraordinaire;][]{Jarrett2007}.

The SMGPS flux errors were calculated using a method similar to \cite{Ramatsoku2016}: by determining the variation of fluxes measured in four emission-free areas surrounding the source, over the same extent as the source. 
Two outliers (HIZOA J1542-55 and J1605-51; open circles) are not used because their HIZOA baselines show wiggles with high noisy spikes which might have affected the width and height determination of the profiles. The red dashed line
is the result of fitting the data points (excluding the two outliers)
with a linear regression least-squares fit algorithm. This routine returns a near-perfect slope of 1.03 ±\,0.06, with an R$^2$ value of 0.92\,--\,demonstrating strong agreement between SMGPS and HIZOA fluxes. 
The ATCA data of the two GLIMPSE galaxies provide the only interferometric counterparts, and their agreement in flux is reassuring. 

We furthermore checked consistency in positions, velocities, and linewidths. 
The relatively high mean angular separation of $2\farcm6\pm2\farcm0$\ is primarily caused by HIZOA's resolution (15\farcm5). A better positional accuracy indicator is the offset with the GLIMPSE galaxies, which ranges between a mere 6--$\ang{;;8}$. The shapes of the \HI\ profiles (HIZOA vs SMGPS) are found to be very similar as well as the resulting \HI~parameters. 
The SMGPS values of $V_{\rm sys}$, \Wtwenty~and \Wfifty~show no deviation from linearity compared to HIZOA. The SMGPS values have a mean error of 14, 20 and 30\,\kms\ respectively, and a standard deviation below the channel width (11, 30, 42\,\kms). The \HI\ parameters of the two GLIMPSE galaxies fall into the same range, with mean offsets of 23\,\kms, 12\,\kms and 15\,\kms. This confirms the SMGPS calibration and \HI\ parameters to be of good quality, well within expectations given the SMGPS channel width of 44\,\kms.

\subsection{Large-Scale Structure} 
\label{section:LSS}

\begin{figure} 
    \centering 
    \includegraphics[width=0.4\textwidth]{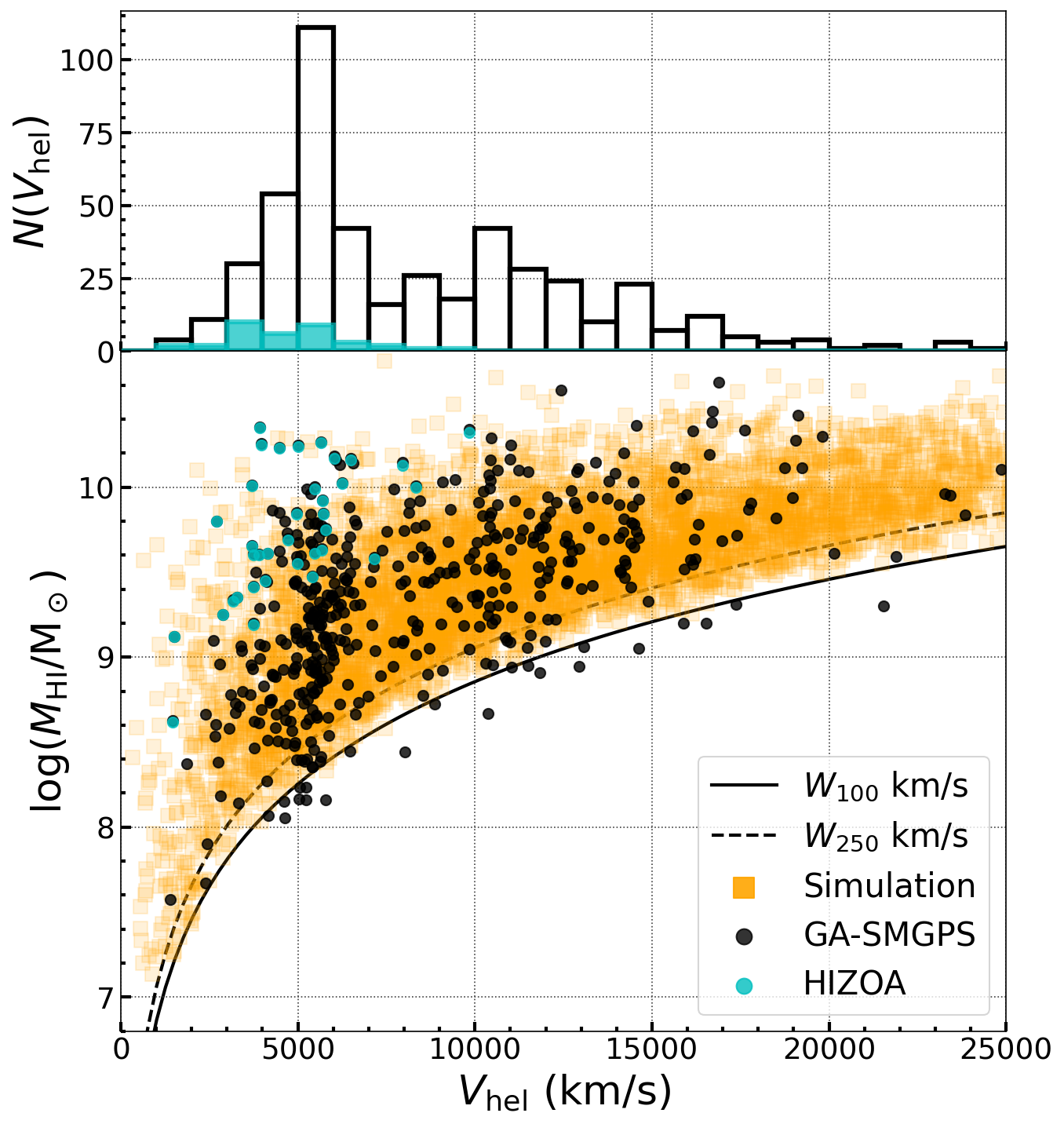}
    \caption{\HI-masses as a function of heliocentric velocity ($\vhel$) for the GA-SMGPS detections (black dots), with the HIZOA counterparts in cyan. The dashed and solid curves represent 5-$\sigma$ detection limits for \Wfifty\ of 100 and 250\,\kms, respectively. Simulations of the full SMGPS survey area (528\,deg$^2$) and the GA-SMGPS specifications are displayed in orange. The velocity distribution is given at the top, with HIZOA galaxies in cyan.}
    \label{fig:sense_curve_sim}
\end{figure}

As a first step in our examination of the unveiled large-scale structures, as well as the \HI-mass detectability of the GA-SMGPS, we display in Fig.~\ref{fig:sense_curve_sim} the \HI-mass as a function of recessional velocity. 
Our detections are superimposed on a simulation that was derived following the precepts given in \cite{Staveley-Smith-Oosterloo-2015} for the SKA \HI~science case, tuned to the specifications of the SMGPS: 0.45\,mJy rms, 44\,\kms\ channels and the full survey area of 528~deg$^2$.
The simulation assumes the HIPASS non-evolving 2D \HI\ mass-velocity width function derived in \cite{Zwaan2005}, placing galaxies at random out to the corresponding sensitivity limit of the observations. Loss of S/N ratio due to velocity width and telescope resolution are both taken into account \citep[see also][]{duffy2012}.
These predictions are shown as orange squares, the GA-SMGPS detections as black dots, and the galaxies in common with HIZOA in cyan. The two curves represent 5-$\sigma$ detection limits for galaxies with $\Wfifty = 100$\,\kms (dwarfs) and 250\,\kms (normal spirals) for an rms of 0.45~mJy. 

Figure~\ref{fig:sense_curve_sim} demonstrates that the \HI-detections reach the limits of the 5-$\sigma$ sensitivity curves throughout the survey volume, and that we are sensitive to $M^{*}_{\rm HI}$ galaxies out to its edge. The \HI-mass ranges between \MHIunit\ : 7.6--10.6. The distribution is far from uniform, revealing distinct clustering within the survey. A striking overdensity hovers around the GA distance range ($\vhel\sim5000$\,\kms) where it probes the gas-rich galaxy population over an order of magnitude deeper than HIZOA. 
The velocity histogram of this wall-like structure shows broad shoulders reaching from about 3500--6500\,\kms (see also Fig.~\ref{fig:simulation_hist}), a typical signature of superclusters \citep[e.g.,][]{RKK2017}. Another peak around 10500\,\kms is not associated with any known structure. A small rise in detections around 14500\,\kms\ supports an earlier proposed connection between the rich Ara and Triangulum-Australis clusters below the Galactic plane \citep[$\ell,b,V = 329\fdg3,-9\fdg9,14634\,\kms$ and $324\fdg5,-11\fdg6,15060\,\kms$; e.g.,][]{woudt2004,Radburn-Smith2006} and the Shapley SCL \citep{Proust2006}.

\subsection{The GA Wall across the ZoA}
\begin{figure}
    \centering
    \includegraphics[width=0.43\textwidth]{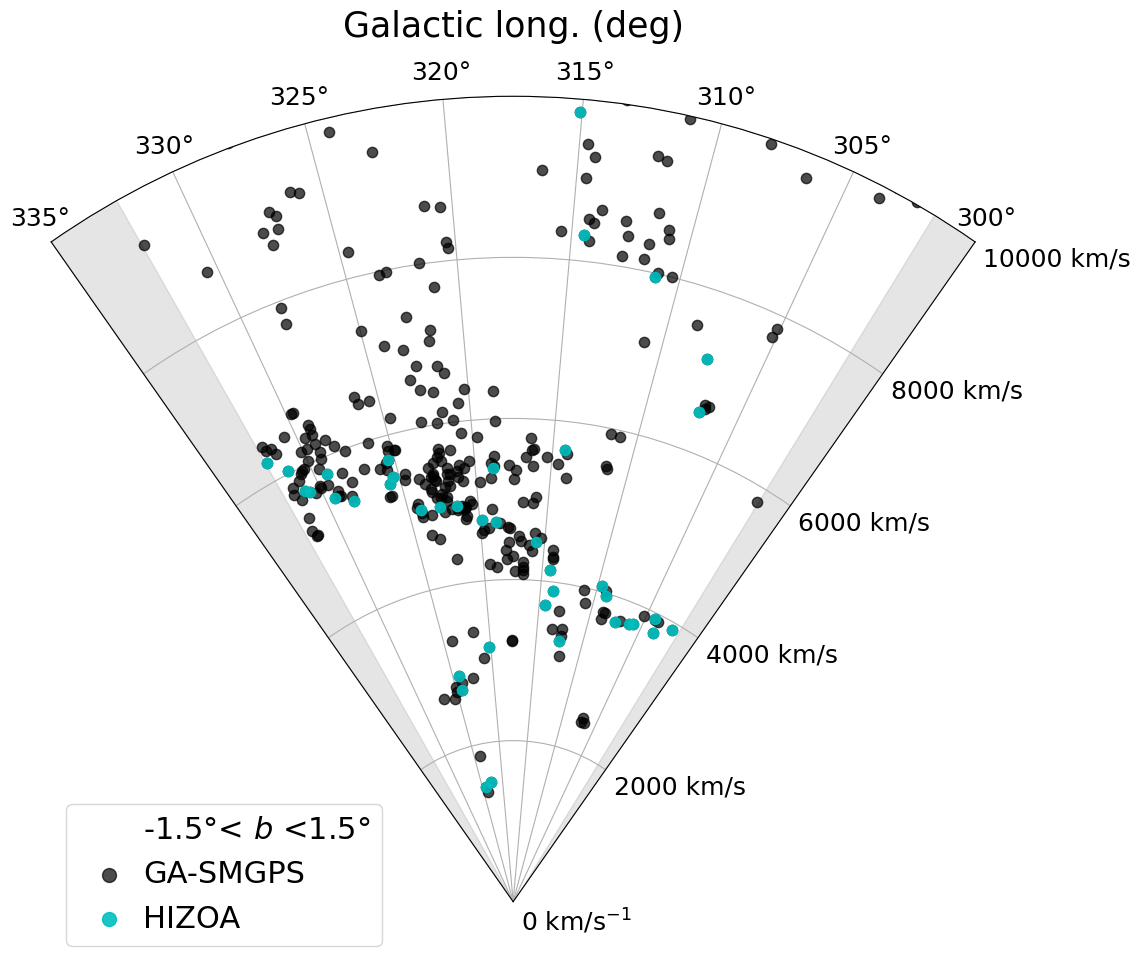}
    \caption{Wedge diagram of GA-SMGPS galaxy detections (black dots) and the HIZOA galaxies (cyan dots) for $V_{\rm hel}\,\leq$\,10000\,\kms.}
    \label{fig:GA_GPS_wedge}
\end{figure}

To investigate the structure and morphology of the GA crossing in more detail, Fig.~\ref{fig:GA_GPS_wedge} displays the GA-SMGPS in a redshift wedge limited to $\vhel$<10000\,\kms. The GA Wall is well-defined, starting from just below 4000\,\kms\ at the lower longitudes, up to about 6500\,\kms\ on the other side. It clearly is {\em the} dominant feature in this narrow cone. The massive HIZOA spirals (cyan dots) follow the shape outlined by the GA-SMGPS closely, but the latter shows a much denser and broader morphology with its \HI\ population of galaxies that reach well into the dwarf regime. 

\begin{figure}
    \centering
    \includegraphics[width=0.48\textwidth]{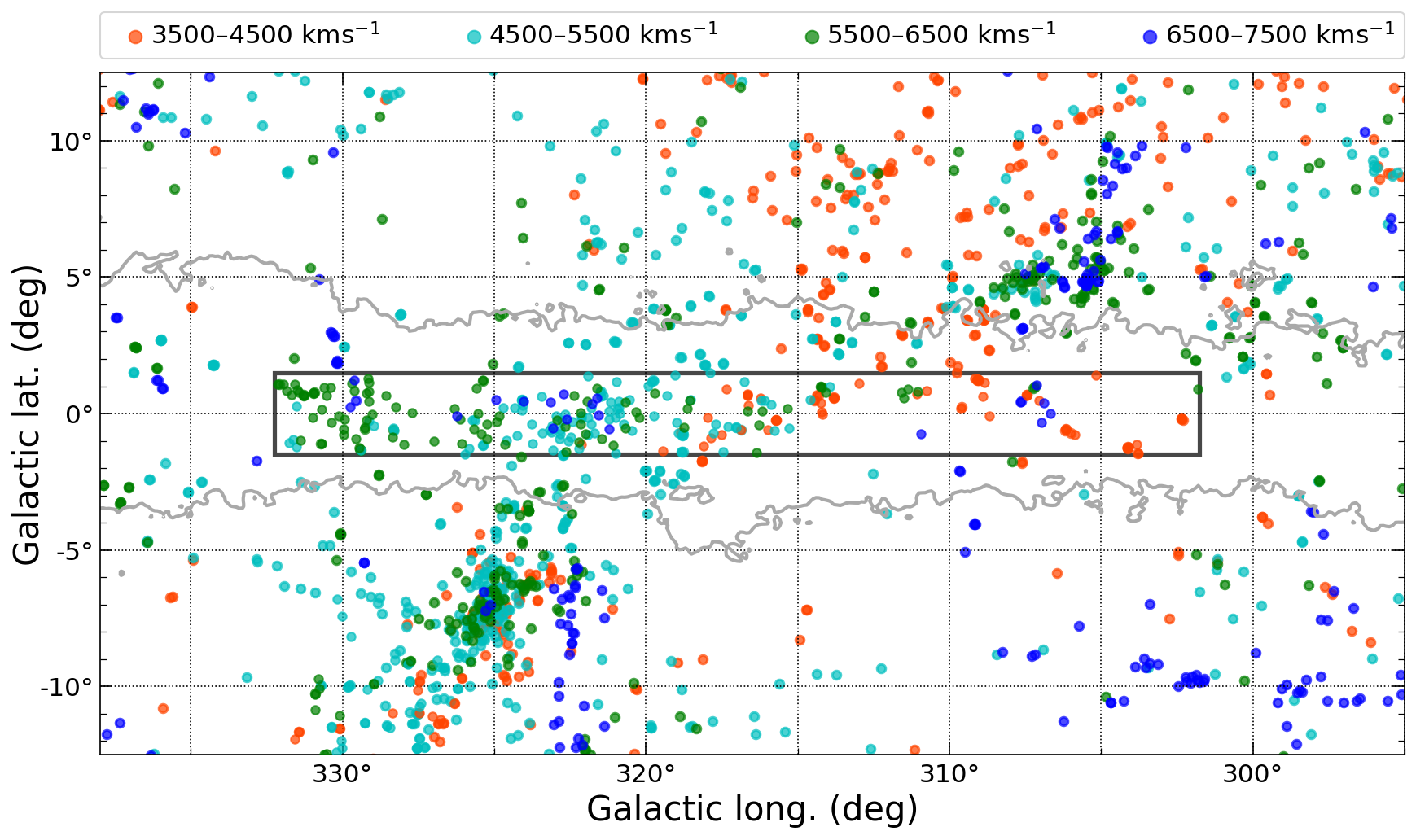}
    \caption{On-sky distribution of galaxies from GA-SMGPS, HIZOA, HyperLeda and WALLABY for the velocity range $3500 < \vhel < 7500$\,\kms. The black box represents the GA-SMGPS search area. The grey contour marks the $A_{\rm B} = 3\fm0$ IRAS/DIRBE extinction level.} 
    \label{fig:LEDA_zoom}
\end{figure}

For more insight into the distribution of the GA connection, Fig.~\ref{fig:LEDA_zoom} displays an on-sky plot of the GA-SMGPS in its wider surroundings ($295\degr<\ell<340\degr; |b| < 12\fdg5$) including galaxies from HIZOA ($|b|<5\degr$), \href{http://leda.univ-lyon1.fr/}{HyperLeda} and the WALLABY pilot survey field in Norma \citep{westmeier2022}. The colour coding subdivides the velocities into bins of $\Delta V = 1000$\,\kms\ over the velocity range 3500--7500\,\kms. The DIRBE/IRAS extinction contour \citep[e.g.,][]{Schlafly2011} of $A_{\rm B} = 3\fm0$ is added to indicate where the ZoA becomes completely opaque to optical galaxies \citep{RKK-Rv2000,Woudt&RKK2001}.

The Norma cluster at $(\ell,b$) = ($325\fdg3,-7\fdg2$) is the most prominent structure in Fig.~\ref{fig:LEDA_zoom}. It appears to be {\em the} centre of a web from which a broad wall extends into the GA-SMGPS around $\ell \sim 322\degr$ (see green, cyan and blue dots). At the same longitude and velocity range, the concentration at higher latitudes seems offset from the main GA Wall in this figure, contrary to what is seen in Fig.~\ref{fig:GA_GPS_wedge}. This might suggest that the overdensity is related to a wall that connects to the Ophiuchus cluster \citep[$\ell,b,V\approx0\fdg5, 9\fdg5, 8500$\,\kms;][]{Wakamatsu2005}, a hint of which had been seen in the HIZOA Galactic Bulge extension \citep{rkk2008} and also in the SMGPS LV data \citep{sushma_in_prep}. A narrow filamentary connection seems to emanate from the Norma cluster (red and green dots), extending towards the Centaurus-Crux cluster at $306\degr,6\degr,6200\,\kms$ \citep{Nagayama2006,Radburn-Smith2006,Kocevski2007}; while the red GA-SMGPS dots around  $\ell: 304\degr-310\degr$ may well be associated with the PKS1343 cluster \citep[$309\fdg7, 1\fdg7,~3872$\,\kms;][]{Nagayama2004,schroeder2007}.

\begin{figure} 
    \centering 
    \includegraphics[width=0.45\textwidth]{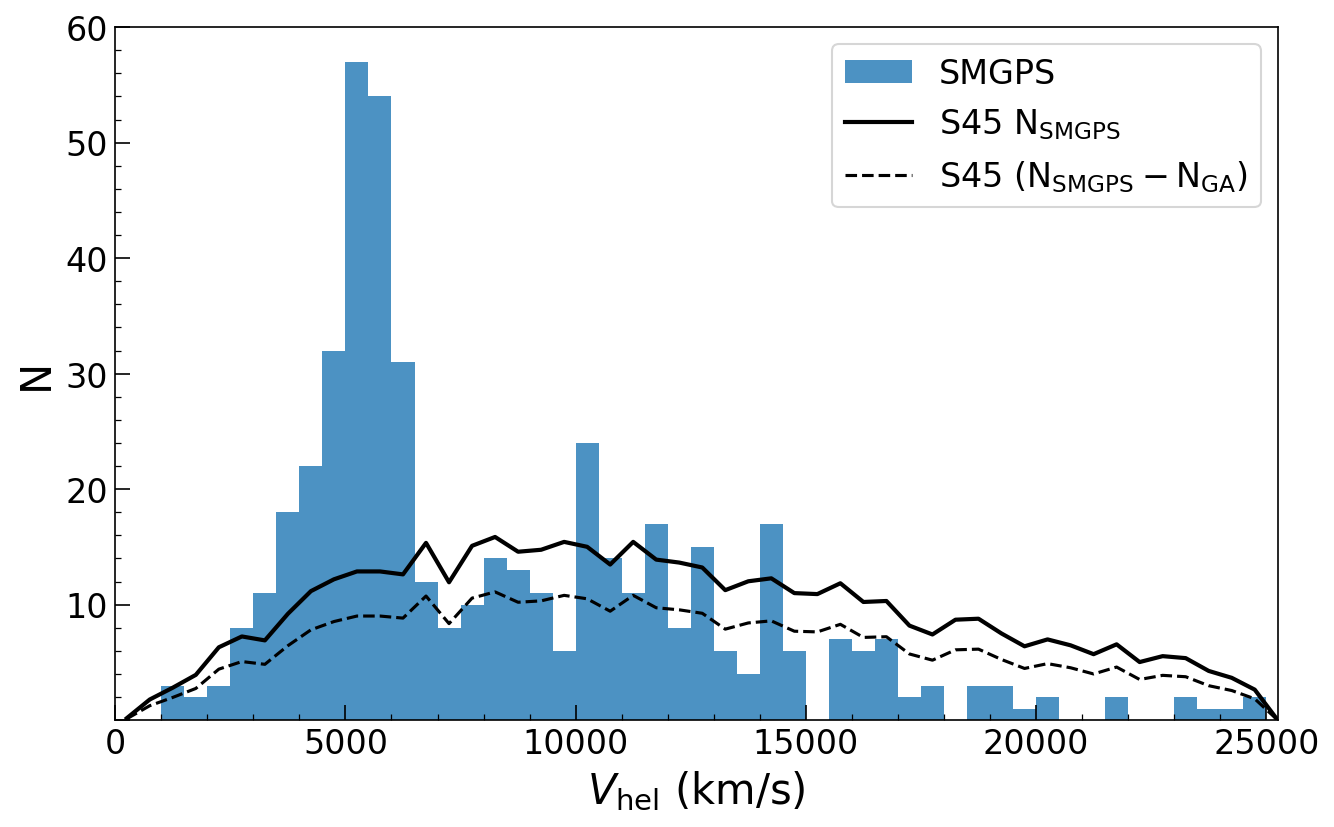}
    \caption{The velocity distribution of the GA-SMGPS galaxy candidates (blue histogram). The solid black gives the S45 predictions scaled to the total GA-SMGPS detections; the dashed line represents S45, \emph{excluding} the GA-overdensity above the S45 prediction for 3500--6500\,\kms.}
    \label{fig:simulation_hist}
\end{figure}

As a final step, Fig.~\ref{fig:simulation_hist} compares the GA-SMGPS velocity distribution to counts predicted by the simulation (henceforth~S45) mentioned in Section~\ref{section:LSS}. 
The blue histogram depicts the GA-SMGPS results, and the black solid line the S45 predictions. Note that we scaled the S45 simulations to the GA-SMGPS detection counts (not the survey area) because of the noise variations in GA-SMGPS. Although the mean rms of the mosaics (0.47 mJy) is close to the 0.45 mJy in S45, its range extends to values of 0.53 mJy. 
Moreover, all mosaics suffer varying degrees of noise fluctuations localised around continuum residuals. The S45 nevertheless provides a good indication of the expected counts as a function of redshift, and overall is a good representative of a more homogeneous galaxy distribution, given its nearly six times larger survey area compared to GA-SMGPS (528\,deg$^2$ versus 90\,deg$^2$).

The difference between the S45 and GA-SMGPS detections highlights the enormity of the GA Wall overdensity: 45\% of our detections out to 25000\,\kms\ lie within $3500 - 6500$\,\kms, while S45 predict this to be around 15\%, hence a factor of $f$\,=\,3 lower. This would imply an overdensity of $\Delta = 2$ at the GA distance range volume. 
One could argue that the galaxies conforming the GA overdensity, i.e., the number of galaxies above the solid line (S45), should be subtracted when scaling S45\,--\,to better reflect a uniform galaxy distribution in this volume. 
If such a re-scaling were applied (dashed line in Fig.~\ref{fig:simulation_hist}), the GA Wall would obviously be even more pronounced ($f$$\simeq$\,4.3, leading to an overdensity of $\Delta$$\simeq$\,3.3). 

At distances beyond $\vhel > 15000$\,\kms, the counts consistently lie below the S45 curves. This could indicate lower completeness due to noise variations, which could affect the high distance range more strongly. But this is not corroborated by results obtained in other parts of the SMGPS, which show no systematic decrease at the higher redshifts. Moreover, the highest redshift detections in GA-SMGPS all lie close or even below the \HI-mass sensitivity curves, even the $\Wfifty = 250$\,\kms\ limit\,--\,a value typical for galaxies in that \HI-mass range. The low counts therefore seem more likely to mark a real underdensity, possibly the signature of an extended distant void.

\section{Conclusions}
\label{section:GPS_conclusion}

The deep southern SARAO MeerKAT Galactic Plane survey ($|b| \la 1\fdg5$) was analysed over an area of $302\degr\leq\ell\leq332\degr$ to chart the GA Wall across the ZoA using the 21\,cm emission line of galaxies. The data comprising 157 fields were mosaicked into 10 \HI\ data cubes to which the automated source finder and parameterisation algorithm SoFiA \citep{Serra2015} was applied. Its parameter settings were optimised following a deep visual inspection of one mosaic.

Overall, 477 galaxy candidates were identified in GA-SMGPS. The considerably higher sensitivity of the SMGPS compared to HIZOA led to a substantial increase in detections: 382 versus 42 within the HIZOA volume ($\vhel \leq 12000$\,\kms). A quality assessment of the SMGPS and HIZOA found no deviations of the integrated flux over  the probed flux range of 2--30\,Jy\,\kms. In view of the SMGPS channel widths ($\sim 44$\,\kms), the velocities and linewidths agree very well (below one channel). 
\HI\ detections were found down to a limit of \MHIunit\ = 9.8 to the edge of the full survey volume, and down to 8.5 at the GA distance. 

The structures outlined by GA-SMGPS clearly confirm the continuation of the GA Wall across the innermost ZoA. The prominence and detail of the Wall are striking: the overall galaxy counts within the velocity range 3500--6500\,\kms\ are a factor of $f$$\sim$3--5 higher compared to S45, and while the GA-SMGPS wall is wider and more densely populated, it closely follows the `skeleton' structure outlined by HIZOA\,--\,consisting largely of \HI-massive spirals. At the largest longitudes, the Wall seems to extend to higher velocities, suggestive of a connection with the Ophiuchus SCL \citep[e.g.,][]{sushma_in_prep, Wakamatsu2005}.  
 
The highly encouraging results of the SMGPS clearly demonstrated that deep interferometric \HI~surveys have the power to uncover the large-scale structure of galaxies, even at the lowest, most optically opaque latitudes. The deep \HI~observations of SMGPS provide a clear view of the GA ZoA crossing: it appears as a smooth large wall-like structure with inter-connecting filaments. 

Our ultimate goal is to exploit the entire SMGPS for hidden large-scale structures. Analysis in two other dynamically important regions, the Vela SCL (Rajohnson et al., in prep.) and the LV \citep{sushma_in_prep} are close to completion. The final papers in these series will be dedicated to a thorough completeness analysis and the derivation of the \HI-mass function, as well as its dependence on environment.


\section*{Acknowledgements}

We would like to thank J.M. van der Hulst and T. Westmeier for their valuable input.
This research was supported by the South African Research Chairs Initiative (SARChI) of the Department of Science and Technology and National Research Foundation. 
The MeerKAT telescope is operated by the South African Radio Astronomy Observatory, which is a facility of the National Research Foundation, an agency of the Department of Science and Innovation. 
This project has received funding from the European Research Council (ERC) under the European Union’s Horizon 2020 research and innovation programme (grant agreement no. 679627). 


\section*{Data Availability}

The full GA-SMGPS galaxy catalog is presented in \hyperref[appendixb]{Appendix B}, and will be submitted to SIMBAD. 
Upon formal publication, the full catalog will be made available as online supplementary material, as well as an atlas of all galaxy candidates. 



\bibliographystyle{mnras.bst}
\bibliography{GA} 


\clearpage
\appendix


\appendix
\setcounter{table}{0}
\onecolumn

\hfill
\phantomsection
\section*{Appendix A}
\label{appendixa}
\hfill \break

\noindent\textbf{SoFiA v2.3.1 Parameter Settings}
\hfill \break

\fontfamily{cmtt}\selectfont
\footnotesize

\noindent \# Global settings \\
pipeline.verbose           =  false \\
pipeline.pedantic          =  false \\
\\
\# Input \\
input.data                 =  /idia/projects/vela/V1\_GA\_CARACal/output\_T19/mosaics/T19\_gal.fits \\
input.region               =  1457,5659,355,3959,9,570 \#square region of 3x3 deg (slightly different for different mosaics) \\
input.gain                 =   \\
input.noise                =  /idia/projects/vela/V1\_GA\_CARACal/output\_T19/mosaics/T19\_gal\_noise.fits  \\
input.weights              =   \\
input.mask                 =   \\
input.invert               =  false \\
\\
\# Flagging \\
flag.region                =   \\
flag.auto                  =  false \\
flag.threshold             =  5.0 \\
flag.log                   =  true \\
\\
\# Noise scaling \\
scaleNoise.enable          =  true \\
scaleNoise.mode            =  spectral \\
scaleNoise.gridXY          =  0 \\
scaleNoise.gridZ           =  0 \\
scaleNoise.interpolate     =  false \\
scaleNoise.scfind          =  true \# normalises the noise after each S+C smoothing step \\
\\
\# S+C finder (smooth \& clip) \\
scfind.enable              =  true \\
scfind.kernelsXY           =  0, 10, 20, 30 \\
scfind.kernelsZ            =  0, 3, 7 \\
scfind.threshold           =  3.5 \#4.0 \\
scfind.replacement         =  1.5 \\
scfind.statistic           =  mad \\
scfind.fluxRange           =  negative \\
\\
\# Linker \\
linker.radiusXY            =  5 \\
linker.radiusZ             =  1 \\
linker.minSizeXY           =  8 \\
linker.minSizeZ            =  2 \\
linker.maxSizeXY           =  0 \\
linker.maxSizeZ            =  21 \#31 \\
\\
\# Reliability \\
reliability.enable         =  true \\
reliability.threshold      =  0.70 \\
reliability.scaleKernel    =  0.25 \# Sofia 2.4.0 will use the auto-kernel feature if set to 0 \\
reliability.minSNR         =  3 \# Replaced reliability.fmin in latest Sofia2 update \\
reliability.plot           =  true \\
reliability.debug          =  true \\
\\
\# Parameterisation \\
parameter.enable           =  true \\
parameter.wcs              =  true \\
parameter.physical         =  true \\
parameter.prefix           =  SoFiA \\
parameter.offset           =  true \# The position parameters will be relative to the full cube (when specifying input.region) \\
\\
\# Output \\
output.directory           =  /users/nadia/Sofia2/ \\
output.filename            =  T19 \\
output.writeCatASCII       =  true \\
output.writeCatXML         =  false \\
output.writeCatSQL         =  false \\
output.writeNoise          =  false \\
output.writeFiltered       =  false \\
output.writeMask           =  false \\
output.writeMask2d         =  false \\
output.writeMoments        =  false \\
output.writeCubelets       =  false \\
output.marginCubelets      =  0 \\
output.overwrite           =  true

\normalsize
\normalfont

\clearpage


\appendix
\setcounter{table}{0}
\onecolumn

\hfill
\phantomsection
\section*{Appendix B}
\label{appendixb}

\hfill \break
\noindent \textbf{The SARAO MeerKAT Galactic Plane Survey: Full Catalog (GA region)}

\hfill \break
A brief description of the parameters follows. Parameters are given by SoFiA unless otherwise stated.

\hfill \break
(1) SMGPS identifier, reflecting the equatorial coordinates [SMGPS-HI-Jhhmmss\,±\,ddmmss]. \\ 
(2)~Mosaic name~[T10\,--\,T19].  \\ 
(3)~Galactic longitude~[deg]. \\ 
(4)~Galactic latitude~[deg]. \\ 
(5)~Peak flux density\,--\,taken from the \HI~profile~[Jy]. \\ 
(6)~Integrated flux~[Jy\,\kms]. \\ 
(7)~Integrated flux error\,--\,measured in emission-free areas surrounding the source~[Jy\,\kms]. \\ 
(8)~Local rms\,--\,as measured in the corners of the sub-cube~[Jy\,beam$^{-1}$]. \\ 
(9)~Heliocentric velocity (optical convention)~[\kms]. \\ 
(10)~Line width at 20\% of the peak flux density, measured in the observer's frame~[\kms]. \\ 
(11)~Line width at 50\% of the peak flux density, measured in the observer's frame~[\kms]. \\
(12)~\HI~mass~[log\Msun], calculated as follows:
\begin{equation}
    M_{\text{HI}} = \frac{2.356\times 10^5}{1+z} D^2 S_{\text{int}}~,\nonumber
\end{equation}
where $D$ is the approximate luminosity distance in Mpc, calculated by $D = \frac{V_{\rm hel}}{H_0}$, where $H_0$ is taken as 70 \kms Mpc$^{-1}$.
Note \HI~masses have not been corrected to the barycentre of the Local Group ($V_{\rm LG}$), nor corrected for the limited instrumental resolution. \smallskip \\ 
(13)~Flag~--~category 1 (high confidence) or 2 (slightly lower confidence). \\
(14)~Note~--~Comment, or name of counterpart. Names starting with `J' are HIZOA names. Note, names from earlier HIZOA publications (e.g., HIPASS, HIZSS) and dedicated follow-up observations are omitted. 
\\

\noindent Footnotes are found at the end of the table. 
{\footnotesize
\begin{longtable}{c@{\hspace{2pt}}c@{\hspace{2pt}}rrrrrrrrrr@{\hspace{3pt}}c@{\hspace{3pt}}l}
\captionsetup{font=bf}\\
\caption*{The GA-SMGPS galaxy catalog.}\\
\hline \hline
\noalign{\smallskip}
SMGPS name & Mosaic &$\ell$ & $b$& $S_{\text{peak}}$ & $S_{\text{int}}$ & err$_{S_{\text{int}}}$ & RMS & $V_{\text{hel}}$ & \Wtwenty & \Wfifty & log\MHI & Flag & Note or \\
J2000 & TXX & deg & deg & Jy & Jy\,km/s & Jy\,km/s & Jy\,km/s & km/s  & km/s & km/s & log\Msun & 1/2 & counterpart \\
(1) & (2) & (3) & (4) & (5) & (6) & (7) & (8) & (9) & (10) & (11) & (12) & (13) \\
\noalign{\smallskip}
\hline
\noalign{\smallskip}
\endfirsthead
\caption*{GA-SMGPS galaxy catalog --Continued.}\\
\hline\hline
\noalign{\smallskip}
SMGPS name & Mosaic &$\ell$ & $b$& $S_{\text{peak}}$ & $S_{\text{int}}$ & err$_{S_{\text{int}}}$ & RMS & $V_{\text{hel}}$ & \Wtwenty & \Wfifty & log\MHI & Flag & Note or \\
J2000 & TXX & deg & deg & Jy & Jy\,km/s & Jy\,km/s & Jy\,km/s & km/s  & km/s & km/s & log\Msun & 1/2 & counterpart \\
(1) & (2) & (3) & (4) & (5) & (6) & (7) & (8) & (9) & (10) & (11) & (12) & (13) \\
\noalign{\smallskip}
\hline
\noalign{\smallskip}
\endhead
\noalign{\smallskip}
\hline
\endfoot

J124141-620158 & T19 & 301.788 & 0.817  & 0.008 & 1.25  & 0.18 & 0.33 & 12051 & 228 & 212 & 9.93  & 1 &                                           \\
J124160-615621 & T19 & 301.821 & 0.912  & 0.008 & 0.46  & 0.09 & 0.33 & 5817  & 116 & 82  & 8.87  & 1 &                                           \\
J124342-620018 & T19 & 302.024 & 0.853  & 0.005 & 0.42  & 0.06 & 0.29 & 12042 & 277 & 277 & 9.45  & 1 &                                           \\
J124404-624235 & T19 & 302.088 & 0.150  & 0.006 & 0.94  & 0.13 & 0.26 & 10920 & 267 & 225 & 9.72  & 1 &                                           \\
J124431-622415 & T19 & 302.130 & 0.457  & 0.010 & 0.78  & 0.11 & 0.29 & 11780 & 138 & 82  & 9.70  & 1 &                                           \\
J124532-624150 & T19 & 302.255 & 0.167  & 0.010 & 0.84  & 0.12 & 0.26 & 11828 & 126 & 97  & 9.73  & 1 &                                           \\
J124624-630308 & T19 & 302.362 & -0.186 & 0.093 & 30.84 & 3.04 & 0.35 & 3912  & 489 & 455 & 10.35 & 1 & \scriptsize{J1246-63, IRAS 12362-5919}        \\
J124734-630933 & T19 & 302.494 & -0.291 & 0.012 & 1.66  & 0.24 & 0.28 & 10105 & 266 & 241 & 9.90  & 1 &                                           \\
\end{longtable}
}

\textsuperscript{1}Consists of a galaxy A ($V_{\rm hel}\,\simeq\,$16483\,--\,16532\,km/s), which could be a close pair, plus an offset neighbour B ($V_{\rm hel}\,\simeq$\,16928\,km/s). \\
\hspace*{\parindent}\textsuperscript{2}{Special case: very extended on-sky. The dispersed emission does not appear clear in the moment maps.} \\
\hspace*{\parindent}\textsuperscript{3}{Other names: IRAS 15147-5527 and WISEA J151829.66-553849.4}


\clearpage


\bsp	
\label{lastpage}
\end{document}